\documentclass[]{article}
\def\half{{1\over 2}}
\def\ben{\begin{equation}}
\def\een{\end{equation}}
\def\bea{\begin{eqnarray}}
\def\eea{\end{eqnarray}}
\def\p{\partial}
\input amssym.def
\input amssym.tex
\begin{document}

\hfuzz=100pt
\begin{flushright}
MIT-CTP-3557 \\  
DAMTP-2004-123\\
CERN-PH-TH/2004-218\\
hep-th/0411119
\end{flushright}

\begin{center}

\vspace{2cm}

{\LARGE {\bf Matrix Cosmology\footnote{Slightly extended version of the 
 talk given by G. W. Gibbons at the Mitchell Institute Conference on
Cosmology}} }

\vspace{1cm}

Daniel Z. Freedman$^{a,b}$, Gary W. Gibbons$^c$ and Martin Schnabl$^{b,d}$

\vspace{1cm}
$^a$ {\it Department of Mathematics, Massachusetts Institute of Technology, 
Cambridge, MA 02139 USA} \\
$^b$ {\it Center for Theoretical Physics, Massachusetts Institute of 
Technology, Cambridge, MA 02139 USA} \\
$^c${\it DAMTP, Centre for Mathematical Sciences,
 Cambridge University \\ Wilberforce Road, Cambridge CB3 OWA, UK} \\
$^d${\it Theory Division, CERN, CH-1211 Geneva 23, Switzerland}

\vspace{0.5cm}
e-mails: dzf@math.mit.edu, g.w.gibbons@damtp.cam.ac.uk, martin.schnabl@cern.ch

\end{center}

\vspace{1cm}

\begin{abstract}
Some speculative preliminary ideas relating matrix
theory and cosmology are discussed. 
\end{abstract}
\vfill \eject

\pagebreak

\tableofcontents

\section{Motivation}
This is a report on some on-going work in which 
an attempt is made to explore how to incorporate the basic ideas of
cosmology into M-Theory. It may be seen either in the context of much recent
work
on time dependent backgrounds in String Theory,  or in its own right,  as
a speculative approach to cosmology aimed at ultimately 
taking us beyond the standard
Friedman-Lemaitre paradigm. More concretely, our motivations are

\medskip 
\noindent $\bullet$  The BFSS matrix model \cite{BFSS} 
is claimed to provide a fundamental 
quantum mechanical description of ``M-Theory''\footnote{
For a useful review of M(atrix) theory see e.g. \cite{Wati}.}.

\medskip
 \noindent $ \bullet$ It replaces conventional  spacetime concepts, 
such as {\sl commuting} coordinates,
 with inherently non-classical notions such as {\sl non-commuting}
 coordinates.

\medskip
 \noindent $\bullet$ It should, therefore, surely have something
 deep to say about
 the structure of the universe. 

\medskip 
\noindent $\bullet$ In particular one should be able to use 
it to address 
such issues as the existence
and significance of such things as ``The Wave function of the Universe''.

\medskip\noindent 
In this talk we present some {\sl rudimentary} and very {\sl preliminary}
 ideas aimed at understanding how we should think about cosmology in the
 language of matrix theory. It is a report of work done partly
in collaboration with S. Alexander. 
 The only previous work on this topic known to us 
is that of Alvarez and Meessen \cite{Alvarez}.

\section{Matrices and D0-particles}

One may view the matrix model in two slightly different ways.

\medskip \noindent$\bullet$  Following BFSS, as 
the limit $N \rightarrow \infty $ of a 
super-quantum-mechanics of 9 $N \times N$  Hermitian matrices.

\medskip \noindent$\bullet$ Following earlier work by de Wit, Hoppe and 
Nicolai \cite{deWit}, as a regularization
of the super-membrane of 11-dimensional supergravity.

\medskip \noindent$\bullet$ Both approaches lead, because of the
 high amount of super-symmetry, to 10-dimensional 
super-Yang-Mills with gauge group $G=U(N)$ and fermions
in the adjoint representation reduced to one spacetime dimension.
 In Coulomb gauge, one replaces the $\frak {u}(N)$ valued connection one-forms
 $A_\mu({\bf x},t)$ by their 9 spatial components $A_i(t)$ which are
the 9 Hermitian matrices $X^i(t)$ of the model.

\medskip \noindent$\bullet$ From the membrane point of view one passes to
 light-cone gauge and the $X^i$ represent the 9 transverse
 components of the membrane coordinates. The residual bosonic gauge-invariance
consists of ${\rm sdiff}(\Sigma_2)$, area preserving
  diffeomorphism of the membrane 2-manifold
 $\Sigma_2$. The Lie algebra
of  ${\rm sdiff}(\Sigma_2)$ is well known to 
coincide, in some sense at least,
 with $\lim _{N \rightarrow \infty} \frak{u} (N) $.

\medskip \noindent$\bullet$ From the 10-dimensional String Theory 
point of view one should regard
the $X^i$ as representing the 9  non-commuting position coordinates of
$N$ D0-branes, the locations of the ends of fundamental strings.  


\medskip \noindent $\bullet${\it Clusters} of large numbers of D0-particles are
described by classical solutions of
 10-dimensional Type IIA supergravity theory.

\medskip \noindent The BPS states correspond to electrically charged
singular extreme ``black hole'' hole
 solutions where the electric Ramond-Ramond charge couples to a   
gravi-photon field of the 10-dimensional Type IIA super-gravity theory
which may be obtained by dimensional reduction of the 11-dimensional
super-gravity theory. 

The classical solutions, which describe $k$ separated  clusters in force 
balance may be lifted to 11-dimensions 
where they have the structure of $k$ singular vacuum pp-wave solutions
moving parallel to each other. 
The pp-waves are sometimes described as 11-dimensional gravitons, but this
is not really accurate, because even in 11-dimensions,
the solutions have distributional sources.
A better description is as lightlike cylinders 
extending along the 10'th spatial dimension.

\section{Classical Matrix Theory} 

One approach to Matrix Cosmology is via the classical equations
of motion \cite{Alvarez}. We shall briefly describe this, since
it was the original approach that we adopted,
 but later we will,
for reasons to be explained, abandon it
for a rather different picture.     

The basic classical equations of matrix theory are
\ben
\fbox{$ \displaystyle {d^2 X ^i \over dt^2} + [X^j, [X^j, X^i]] =
 \lambda X^i,  
$}\een 
where $X^i$ are $n$ $N \times N$ hermitian matrices,
the index $j$ is summed over and $\lambda = {\Lambda c^2 \over 3}$
 is a possible cosmological or mass term \cite{Hoppe,Alvarez,Gao,Patricot}. 

In the BFSS model $n=9$, and one is looking at $N$ D0-branes
 but these equations have been studied more widely as a reduction
of $U(N)$, or, if they are taken traceless, $SU(N)$,  Yang-Mills theory to
one time dimension. As such, there is some evidence for chaotic behaviour.

\medskip \noindent Note that 

\medskip \noindent $\bullet$ If $\lambda =0$, we have a Galilei
 invariant system

\medskip \noindent $\bullet$ If $\lambda \ne0$, we have invariance
under one of the two Newton-Hooke groups, of the two non-relativistic
contractions of the De-Sitter ($\lambda >0$)  or Anti-de-Sitter ($\lambda <0$) 
groups. A description of these groups and their transformation rules 
together with an account of their significance for Newtonian cosmology
with a cosmological constant are given in  \cite{Patricot}. 
 
\medskip \noindent $\bullet $ If one thinks in terms of a mass term, 
then a positive mass squared corresponds
to negative cosmological constant and a tachyonic mass term to
 a positive cosmological constant.

\medskip \noindent $\bullet$ In the case of the BFSS model, 
the equations of motion must be supplemented
by a {\it constraint} on the initial conditions which arises from the Gauss
constraint of the gauge theory
\ben
[ \dot X ^i , X^j ]=0.
\een

\section{Newtonian Cosmology}

Since our equations lack manifest covariance, the most helpful
analogy is with elementary Newtonian Cosmology in 3 spatial dimensions
The brief  presentation of Newtonian cosmology which follows may be
unfamiliar, but it is completely equivalent  to more conventional 
accounts in the literature.  The generalization to
other space dimensions, and indeed to other force laws, is trivial.
 
Newton's equations of motion
for $k$ gravitating particles are   
\ben
\fbox{$ \displaystyle m_a \ddot {\bf r}_a = 
\sum G {m_am_b ({\bf r}_b-{\bf r}_a)
 \over |{\bf r}_a-{\bf r}_b | ^3 }+ \lambda m_a {\bf r}_a. $}
\label{inertia}\een

\noindent $\bullet$ As with the matrix model, 
we either have Galileo invariance
 ($\lambda=0$)
if the cosmological constant vanishes, or if it does not we have Newton-Hooke
invariance.

\medskip \noindent $\bullet$ In order to incorporate the 
{\it Cosmological Principle} we make a {\it Homothetic Ansatz}
\ben
{\bf r}_a (t) =a(t) {\bf x}_a ,
\een  
where the so-called {\it co-moving coordinates} ${\bf x}_a$ 
are independent of time. 

The homothetic ansatz leads to two conditions.

\medskip \noindent $\bullet$ {\it Raychaudhuri's Equation}
\ben
{ \ddot a \over a} = -{\mu \over 3 a^3 } +\lambda . 
\een
 This is the usual equation of motion for
 the scale factor of an expanding universe with a cosmological term and 
pressure free fluid. In what follows we set $\lambda=0$ for
simplicity. It is easy to adapt the discussion to the case $\lambda
\ne 0$
(see \cite{Patricot}).

\medskip \noindent $\bullet$ The co-moving coordinates  must constitute
a {\it Central Configuration}, i.e. a solution of 
\ben
{\mu \over 3} m_a {\bf x}_a + \sum G {m_a m_b ( {\bf x}_b-{\bf x}_a)
 \over |{\bf x}_a-{\bf x}_b| ^3 }=0\,.
\een  

\medskip \noindent The quantity $\mu$ is a constant.
Central configurations are extrema of an auxiliary potential
\ben
\sum { \mu \over 6} m_a {\bf x} _a^2 + \sum \sum G
 { m_a m_b \over |{\bf x}_a -{\bf x_b }| }.
\een  

Recently, with Battye and Sutcliffe \cite{Battye},
one of us  has carried out an extensive
numerical investigation of central configurations which are  minima
or {\it ground-states} of this potential for  
 up to $10 ^4$ particles.
In the case of equal masses $m_a=m$, $\forall a$,
the conclusion is that the minima correspond to  a spherical ball of
particles of uniform density $\mu \over 4 \pi G$.
In other words, if $N$ is the total number of particles
and the radius $\rho$ is defined by 
\ben
{G Nm \over \rho^2}   = { \mu \rho  \over 3}\, ,\label{force} 
\een
then one finds a uniform density of particles  inside the radius $\rho$
and almost no particles outside that radius. The interpretation
of (\ref{force}) should be clear. It is well known that
a spherical shell of matter exerts no force on particles inside it
but an attraction on particles outside given by the total mass 
of the shell. 
The left hand side of (\ref{force}) is the Newtonian
attraction due to the total mass $N m$ interior to radius $\rho$ 
on the thin shell of particles at $\rho$.
The right hand side is the repulsive pseudo-cosmological  force on the thin 
shell which is proportional to the distance $\rho$. We use the term  
`pseudo-cosmological'  to alert the reader to the fact that
we get such a term even if the cosmological constant $\lambda =0$.
It really arises  from the inertial term in the Newtonian equation of
motion (\ref{inertia}). Note that the proper radius of our ball 
is time-dependent and given 
\ben
|{\bf r}|= R= a(t) \rho \, . \label{expansion}
\een

 A ball of uniform density 
 is exactly what one expects on the basis of the usual
 pressure free fluid model. Thus this, slightly unconventional,
approach to Newtonian Cosmology reproduces all of the standard features
 without making  arbitrary assumptions about
fluids, rather these assumptions are {\sl derived} from the model.

\subsection{Quantum Newtonian Cosmology} 

In order to prepare ourselves for Quantum  Matrix cosmology, it may be
worth pausing to 
recall  that  
one can obviously  construct a Wave Function for Newtonian Cosmology
in the framework of non-relativistic quantum mechanics. 
This may not often  be done in discussions of Quantum Cosmology
but it is  entirely straightforward and elementary.   
All that one  needs is  a solution of the multi-particle 
Sch\"rodinger equation  
\ben
i \hbar {\partial  \Psi \over \partial t}= \sum { -{\hbar ^2 \over 2m_a}}
  \nabla _a^2 \Psi  + V \Psi\,,
\een
with $\Psi=\Psi({\bf r}_a)$,  
$\nabla ^2 _a = {\partial ^2 \over   \partial {\bf r}_a^2}$ and  
\ben
V=-\sum \sum G
 { m_a m_b \over |{\bf r}_a -{\bf r}_b| }\,.
\een

\subsection{WKB Approximation}
 
We consider a potential $V=V({\bf r}_a)$ which is homogeneous
of degree $n$. For the case of Newtonian gravity without a cosmological term
corresponds to $n=-1$.
The equation of motion is
\ben
m_a {\ddot {\bf r}}_a= -{\p V \over \p {\bf r}_a }.
\een
The homothetic ansatz is 
\ben \label{homo}
{\bf r}_a =a(t) {\bf x}_a,
\een
where the co-moving coordinates ${\bf x}_a$ constitute
a {\it central configuration}  satisfying
\ben
{ \mu \over 3}  m_a {\bf x}_a = + {\p V \over \p {\bf x}_a} \label{central}
\een
and the scale factor satisfies the Raychaudhuri type equation
\ben \label{ray1}
a^{1-n} {\ddot a}= -{ \mu \over 3}.   
\een 
with first integral or  Friedmann equation 
\ben
{1 \over 2} {\dot a} ^2 + { 1 \over n} {{\mu \over 3}  a ^ n}= { k \over 2}, 
\een
 where $k$ is a constant.
 Taking the dot product of ${\bf x}_a $ with (\ref{central})
and using  Euler's theorem gives the Virial Theorem
\ben
{\mu \over 3}  \sum m_a {\bf x}^2_a = n V({\bf x}_a). 
\een  

The conserved energy is 
\ben
H= { k \over 2}  \sum m_a {\bf x}_a ^2.  
\een

At the JWKB level, the wave function  is
\ben
\Psi \approx e^{{i S \over \hbar}}, 
\een
where $S$ is the relevant solution of the  Hamilton-Jacobi
equation
\ben
\sum { 1 \over 2 m_a} \left({\p S \over \p {\bf r} _a}\right) ^2 + V({\bf r}_a)
= -{\p S \over \p t }.
\een 
In our case the relevant solution is 
\ben
S= \sum  {\dot a \over a} { m_a \over 2}  {\bf r}_a ^2=
 \sum  {a \dot a } { m_a \over 2}  {\bf x}_a ^2 = a \dot a  { H
 \over k}  . 
\een

\subsection{Hartree-Fock approximation}
Here we  suppose all masses equal $m_a=m$ and replace the full wave function
$\Psi({\bf r}_a)$  by the product   
\ben
\Psi({\bf r}_a) \propto  \prod _a  \Psi ^\prime ({\bf r}_a),
\een
where $\Psi^\prime ({\bf r})$ satisfies 

\ben \label{sch}
i \hbar {\partial  \Psi^\prime   \over \partial t}=  { -{\hbar ^2 \over 2m}}
  \nabla ^2 \Psi^\prime    + m U\Psi ^\prime \,,
\een
with $\Psi^\prime =\Psi ^\prime ({\bf r},t)$,  
$\nabla ^2  = {\partial ^2 \over   \partial {\bf r}^2 }$ and  
\ben
\nabla ^2 U= 4 \pi Gm|\Psi ^\prime|^2 \,.
\een
The time-independent Schr\"odinger equation coupled to Poisson's equation
has been studied in a different context 
 where it is referred to as the Schr\"odinger-Newton equation \cite{Tod1}. 

In our case, we  assume that  
\ben
U= F(t) {\bf r}^2 \,,\qquad \Psi^\prime =A(t) e^{iS \over \hbar}, 
\een
We  find from the Poisson's equation that
\ben
A^2 ={3 F \over 2 \pi G m}.
\een

One readily sees that one must have
\ben
S=B(t) {\bf r}^2\,, 
\een
with
\ben
- {3 B \over m}={\dot A \over A}.
\een

Moreover if $A= a^ {-{3 \over 2}}$, then
the scale factor $a(t)$  satisfies the Raychaudhuri equation 
 \ben \label{ray2}
\ddot a = -{4 \pi G m \over 3 a^2 }\,. 
\een
The action is given 
\ben
S= { 1 \over 2} m {\dot a \over a} {\bf r} ^2\,, 
\een
and the wave function by
\ben \label{psihf}
\fbox{$\displaystyle \Psi
 ^\prime \propto { 1 \over a^{3 \over 2}} e^{{ i\over 2 \hbar} 
 m { \dot a \over a} {\bf r} ^2 }\,.  
$}
\een
In this case the Hartree-Fock approximation gives a version of the 
WKB wave function corrected by the prefactor ${1 \over a^{3 \over 2}}$.  

\subsection{Normalization and Energy}
In order to model the central configurations described earlier 
which have a finite number of particles,
we need to use a normalizable wave function. We use the
Hartree-Fock wave function $\Psi^\prime ({\bf r})$ of
(\ref{psihf}) in the region $|{\bf r}| < R$ and take $\Psi ^\prime({\bf r})=0$
for  $|{\bf r}| >R$. This is an exact solution of the
Schrodinger equation (\ref{sch}) within each region, but fails at
the surface $|{\bf r}|=R$. We ignore this issue here.

The normalization integral of $|\Psi^\prime ({\bf r} )|
^2$ has support in a ball of proper radius $|{\bf r}|=R$. The norm
must be time-independent, so we need to take {\it time dependent}
$R(t)$. In fact we need to take  
\ben
R(t)  \propto a(t) \, 
\een
This is consistent with the classical analysis giving Hubble's law
(\ref{expansion}).
 
The classical problem of Newtonian cosmology has a conserved
energy, and we should check the energetics of our
quantum-mechanical model. The energy of a single particle wave
function is
\bea \label{en1}
E &=& \int d^3r \, \left[\frac{\hbar^2}{2m} |\nabla \Psi^\prime|^2\, 
+\,m F(t) {\bf r}^2 |\Psi^\prime|^2\right]
\\
&\propto& \left[\frac{2}{m}B(t)^2 \,+\, F(t)\right] A(t)^2\,\int^{R(t)}\!\! d^3r\,\, {\bf r}^2
\\
&\propto & \half \dot{a}(t)^2 + {2\pi G m \over 3a(t)}.
\eea
The Raychaudhuri equation (\ref{ray2}) has the first integral
\ben 
\half \dot{a}(t)^2 - {4\pi G m\over 3a(t)} = \half k,
\een
so we have
\ben \label{en2}
E \propto \half k + {2 \pi G m \over a(t)}.
\een
This is a constant, as desired, plus a $t$-dependent error term
which we attribute to the sharp cutoff in the wave function. 

Since the quantum-mechanical energy (\ref{en1}) is strictly
positive, we must choose the $k>0$ solution of (\ref{ray2}) with
large $t$ behavior $a(t) \sim k t$. The error term above vanishes
at large $t$. The quantum mechanical model is thus consistent for
an ``open universe''.

\subsection{Wick Rotated Newtonian  Wave function of the Universe}

We have constructed an approximate wave function of a 
simple Newtonian
universe whose WKB approximation gives a  classical solution of
Newton's equations of motion representing an expanding gas of point
particles.  The aim of quantum cosmology is to derive
this wave function, and hence the initial conditions for the universe
from  some more fundamental assumption, such as the No-Boundary Proposal  
of Hartle and Hawking. We shall not dwell on this in detail here but
content ourselves with the following, possibly suggestive, 
 remark. If we take the simplest
(Einstein-de-Sitter)
 solution for the scale factor $a(t) \propto t^{2 \over 3}$, we have
\ben
\Psi ^\prime \propto e^{ {i \over 3 \hbar} { m {\bf r}^2 \over t}}. 
\een    
Curiously, this Euclidean wave function, strictly speaking a solution
of the diffusion equation rather than the Schroedinger,  
 will be normalizable with respect to integrations over the
positions if we Wick rotate, i.e. set
\ben
t=-i \tau,
\een 
with the imaginary time coordinate $\tau$ being real and positive.
One might speculate  that this normalizability of the Wick-rotated  Newtonian 
wave function  is related to  Hartle and Hawking's path integral
approach  to the wave function of the universe.

\section{Homothetic Matrix Cosmology }

After preparing ourselves with a brief  excursion into Newtonian
Cosmology, we return to the matter at hand.  
In matrix  cosmology is it also 
natural to begin by  making  a  homothetic ansatz \cite{Alvarez,Hoppe2}
\ben
X^i= a(t) M^i,
\een
where $M^i$ are independent of $t$.

Substitution leads to 
\ben
{\ddot a \over a^3 } -{\lambda \over a^2} = \mu, \label{ray} 
\een
\ben
\mu M^i + [M^j, [ M^j, M^i]]=0 \label{vac}.
\een

The  idea is to interpret (\ref{ray}) as the analogue of
Raychaudhuri's
equation in cosmology and (\ref{vac}) as the analogue of the
equation governing central configurations in Newtonian cosmology
\cite{Battye} or monopole scattering \cite{Poly}.
Essentially the same equation arises in supersymmetric ${\cal N}= 1^\star$ gauge   
theories  when one is looking for vacua or ground states \cite{Polchinski}.  
For that reason we shall sometimes refer
to solutions of (\ref{vac}) as vacua.  Similar equation also arises in certain solutions 
describing spinning membranes \cite{spinmembr}.  

We begin by looking at Raychaudhuri's equation (\ref{ray}).
It has a first integral
\ben
\left({\dot a \over a}\right)^2 + {k \over a^2} = {\mu a^2 \over 2} +\lambda\,,    
\label{Friedman}
\een
where $k$ is a constant  of integration.
This is the analogue of the Friedman equation in standard cosmology .
As well as a standard cosmological constant or dark energy contribution
given by $\lambda$ we have an exotic matrix contribution to the energy
density
given by
\ben
\rho_M = {3\mu a^2 \over 16 \pi G}\,.
\een
If $\mu$ is positive  this energy density  {\sl increases} as the universe
expands, indicating that the pressure $P_M$ has the opposite sign
to the energy density.
In $d$ spatial dimensions this would lead to 
\ben
P_M=-{d +2 \over d} \rho_M\, . 
\een

From the point of view of matrix theory the most natural
choice for $d$ would be 9. Arguing by analogy with the Newtonian
case one might then regard $a(t)$ as the  scale factor
in Einstein conformal gauge. This then leads to
\ben
P_M=  -{11\over 9} \rho_M\,.
\een  
Later we shall compare this with a model based
on a supergravity solution representing a gas of expanding
D0-branes.

As an example of the general theory, 
consider 3  $N\times N$ matrices  $M^i$, $i=1,2,3$
providing an $N$ dimensional representation of $\frak{su}(2)$,
\ben
[M^1,M^2]=iM^3\qquad {\rm etc}\,.
\een 
This solution actually describes an expanding spherical membrane 
(see \cite{Wati} and references therein). 
From (\ref{vac}) we find
\ben
\mu = -2,
\een
which implies a {\sl negative} energy density and positive pressure.
This looks rather unphysical and so we turn to an anisotropic model.
Recall that
\ben
9=3+3+3,
\een   
and take 3 mutually commuting sets of such matrices, each with its own
scale factor $a(t),b(t),c(t)$ say. If $M^1,M^2,M^3$ are taken to be
diagonal and $\lambda>0$
we shall get exponential expansion for the scale factor $a$ since
\ben
\ddot a= \lambda a\,.
\een
On the other hand, we can have $X^4,X^5,\dots X^9$ oscillating.
In other words
\ben
\fbox{ {\rm 3 directions exponentially expand} } 
\een
\ben
\fbox{ {\rm 6 directions oscillate} } 
\een

This phenomenon is closely related to the well-known {\sl chaotic
  behaviour}
of Yang-Mill reduced to one time dimension and zero space dimensions. 

\subsection{Chaos}

In standard $\frak{su}(2)$ Yang-Mills, one may assume that
the connection is 
\ben
A=X^idx^i=u(t) \tau^1 dx + v(t)\tau^2 dy + w(t) \tau^3 dz \,,
\een
where $\tau^i$ are Pauli matrices. The equations of motion derive from
the Lagrangian
\ben 
L={ 1 \over 2}(\dot u^2 + \dot v^2 + \dot w^2 ) -V(u,v,w),
\een
with
\ben
V(u,v,w)= { 1 \over 2}(u^2v^2 + v^2 w^2 + w^2 u^2 )\,.
\een

The non-negative potential $V$ has {\sl three commutative valleys} along the
three orthogonal   coordinate axes in $u,v,w$ space 
 for which $V$ vanishes.  Studies of the motion
\cite{chaos}  show that the representative particle rattles along
 each valley, eventually returning to the origin and rattling along
another valley. This is rather reminiscent of the behaviour of the
three scale factors $a,b,c$ of a Bianchi IX chaotic cosmology of the
type originally studied by Misner \cite{mixmaster}.  
Introduction of a positive cosmological term, 
(i.e. a tachyonic Higgs mass) leads to the {\sl eventual escape
of the particle in one direction}, provided $\Lambda$ exceeds a
certain threshold.

An obvious extension of this idea is to consider an $SU(2)\times
SU(2)\times SU(2)$ model in which {\sl just three directions expand
exponentially}  and the other  six remain bounded.

\section{D0-particle Cosmology}

According to Type IIA ten-dimensional 
  Supergravity, D0-branes correspond to  extreme black holes, a static
  configuration of  $k$ clusters depending  on a {\sl harmonic  function}
  $H$
on ${\Bbb E}^9$ 
\ben
H= 1+ \sum ^k_{a=1} { \mu_a \over 7 |{\bf x}-{\bf x} _a|}\,,
\een
where $\mu_a$ is quantized, being proportional to $N_a$ for $N_a$ D0-branes
 located at positions ${\bf x}_a$. 

The moduli space is clearly given by $k$ points in ${\Bbb E} ^9$.
The slow motion is governed by a metric induced from the De-Witt
metric
of the Type IIA action. Long ago, Shiraishi showed that this metric is
flat \cite{Shir}.
 In other words, there are no velocity dependent forces quadratic
in velocity. Thus one might anticipate that a cosmology of D0-branes
should expand freely like a {\sl ten-dimensional}  Milne model which
has
\ben
a(t) \propto t\,,  \qquad \rho +9P=0\,.
\een   
This is contradicted by some exact solutions of Type IIA found
by Maki and Shiraishi \cite{Maki}
 some time ago, following earlier work by Kastor
and Traschen. In these one has
\ben
a(t) \propto t^{ 1 \over 9} \,,  \qquad \rho =P\,,
\een
which corresponds to `stiff matter', for which the sound speed is
that of light. Neither of these two equations of state
coincides with that given by the homothetic matrix model
of the previous section. One might try, as was suggested to us
 by Justin Khoury
to `save appearances' by passing to string conformal
gauge, which might seem more appropriate for the homothetic  matrix
model scale factor. However, in this gauge, we would have
\ben
P = -{ 1 \over 3 } \rho,
\een  
which does not coincide with the $P = -{11 \over 9} \rho$
which we obtained from the homothetic matrix model.
In hindsight this is perhaps not so surprising, since the matrix configuration leading 
to this equation of state were rather delocalized and in fact corresponded to extended 
objects. For well localized D-particles with mutually commuting coordinates the quartic 
term in the matrix Lagrangian is effectively zero and therefore the dominant interaction 
between D-particles will come from a 1-loop correction which generates a term $v^4/r^7$ in  
the 2-body Lagrangian. It would be interesting to see whether Newtonian approach to cosmology
based on those interactions could generate Maki-Shiraishi solutions.
We shall not explore that  in detail here
but rather we shall describe the Maki-Shiraishi metrics
and their `hidden supersymmetry'. 
There is no denying that  a cosmology made up entirely of D0-branes, with no
anti-D0-branes might not be thought of as being very realistic.
Nevertheless, the solutions we are about to discuss do exhibit some
extremely interesting features which we hope contains  lessons for 
future, more realistic models.

\subsection{Maki-Shiraishi metrics}
Maki and Shiraishi \cite{Maki}
 considered as  a Lagrangian  in $n+1$ spacetime dimensions
for gravity plus a two-form plus a scalar
 \ben
L=R-{ 4 \over n-1} (\nabla \phi )^2 - e^{ {4 a \over n-1} \phi} F^2 -
 (n-1) e^ {{4 b \over n-1} \phi} \Lambda. 
\een
where $a $ and $b$ (and $\Lambda$) are constants.

They sought solutions of the form
\ben
ds^2 =-H^{-{ 2(n-2) \over n-2+a^2} } dt ^2 + a^2(t) H^{ {2 \over n-2+a^2
  } } d {\bf x} ^2.
\een
\ben
H=1+ { 1 \over a(t) ^{ n-2+{p \over 2} }}  \sum { \mu_a \over  (n-2)
|{\bf x}
  -{\bf x}_a|^{ n-2 }}  \,,
\een
\ben
e^{ {4 a \over n-1} \phi } = a(t)^p H^{-{2 a^2 \over n-2+a^2 }}\,,
\een
\ben
A= \sqrt{ n-1 \over 2(n-2+a^2)} { dt \over a(t) ^{{p\over 2}}} 
\left( 1- {1\over H}\right)\, ,
\een
with $F=dA$.
Maki and Shiraishi found various solutions. The time dependence of the
scale factor depends on the particular solution,
for us the relevant one 
satisfies $n=a^2 =9$, $p=2a^2$, $\Lambda=0$ and
\ben
a(t) = \left({t \over t_0}\right)^{1 \over 9}\,.
\een

If no D0-branes are present the {\sl background metric} is 

\ben
ds_{10} ^2 = -dt ^2 +a^2(t) d{\bf x}^2,
\label{back1}\een
with
\ben
g=e^\phi = t^{ 4 \over 3} \label{back2}.
\een

This is just what one expects for gravity coupled to a
 massless scalar field which behaves just like stiff matter. 
From the point of view of string theory,
 we see that
 we have a time dependent string  coupling constant 
$g$ which increases with
 time from a zero value at the Big Bang.
In other words 
\ben
\fbox{{\rm late times $\longleftrightarrow$ strong coupling } }
\een
\ben
\fbox{{\rm early times 
 $\longleftrightarrow$ weak  coupling } }
\een

This feature remains true if D0-branes are present.
First note that 
one may take $a(t) $ to be constant by making the change  
\ben
t \rightarrow t+ t_0\,,
\een
and letting $t_0 \rightarrow \infty$.
One then obtains the static multi-brane solutions
\ben
ds_{10}^2 =-H^{-{7 \over 8}} dt ^2 + H^{1 \over 8}  d {\bf x} ^2\,,
\een with
\ben
H= 1+ \sum ^k_{a=1} { \mu_a \over 7 |{\bf x}-{\bf x} _a|}\,,
\een
\ben
g=e^\phi = H^{3 \over 4}\,,
\een 
and
\ben
A=\left(1-{ 1 \over H}\right)dt.
\een

Now let us restore the time dependence. One finds, setting $t_0=1$,
\ben
ds_{10}^2 =-H^{-{7 \over 8}} dt ^2 +t^{2 \over 9} 
 H^{ 1 \over 8} d {\bf x} ^2\,,
\een with
\ben
H= 1+ \sum ^k_{a=1} { \mu_a \over 7 t^{ 16 \over 9} |{\bf x}-{\bf x} _a|}\,,
\een
\ben
g=e^\phi = t^{4 \over 3} H^{3 \over 4}\,,
\een 
and
\ben
A=\left(1-{ 1 \over H}\right){dt \over t}\,.
\een
Evidently the general time-dependent solution represents a gas of
D0-branes in 10 dimensions in a background time dependent dilaton field. 
Note that 

\medskip \noindent $\bullet$
 the non-interacting  gas of D0-branes does not  affect the law of
expansion.

\medskip \noindent $\bullet$  
 while the solution at large distances is time-dependent,
with the physical separation of the D0-branes increasing with time,
near  each singularity, i.e. as ${\bf x} \rightarrow {\bf x}_a$, 
 the solution   
is effectively {\sl static}.

\section{Lift to 11 dimensions}
 The Maki-Shiraishi metrics are clearly not supersymmetric,
i.e. they are not BPS because they are time dependent. 
Every Lorentzian spacetime admitting a
Killing spinor field  must admit an everywhere non-spacelike Killing
vector field. However if one lifts the solution to eleven dimensions
using the uplifting formula 
\ben
ds ^2_{11}= e^{-{\phi \over 6}} ds_{10}^2 + e^{ 4 \phi \over 3} (dz +2A)^2,
\een
where $z$ is the eleventh coordinate,
something interesting happens. 

\subsection{The background}
Using the  uplifting formula one finds that the background (\ref{back1},\,\ref{back2}) becomes
\ben
ds _{11}^2 = t^{-{2 \over 9}} (-dt ^2 + t^{2 \over 9} d {\bf x}^2 ) +
t^{16 \over 9} d z^2\,.
\een
If one defines 
\ben
T={9\over 8} t^{8 \over 9}\,,
\een
we get
\ben
ds_{11}= -dT^2 + \left({64 \over 81}\right)^2  T^2 dz ^2 + d{\bf x} ^2 \,.
\een
This is flat space ${\Bbb E} ^9 \times {\Bbb E}^{1,1}$ in
 {\sl Milne coordinates}.
Of course if the tenth coordinate $z$ is taken to be periodic, as it
would be on the M-Theory circle, then  we shall
get the usual orbifold singularities and non-Haussdorf behaviour  
associated  with Misner spacetime \cite{Misner}. 

\subsection{Lifting the general solution} 

We define
\ben
d\tilde t = t^{ 7 \over 9} dt\,\qquad \Rightarrow \tilde t= {9 \over
  16} t^{ 16\over 9} .
\een
We obtain
\ben
ds_{11}^2= d {\bf x} ^2 +t^{16 \over 9} H(dz + 2A)^2 - {dt ^2 \over
  t^{2 \over 9} H}\,.
\een
Now let

\ben
d\tilde z= dz +{dt \over t}\,.
\een
One gets
\ben
ds^2 _{11}=d {\bf x} ^2 + t^{ 16 \over 9} H (d \tilde z) ^2 - 2 d
\tilde z
t^{7\over 9}.
\een

This looks complicated, but
if we define a time independent harmonic function 
\ben
\hat H= { 1 \over 7} \sum {\mu _a \over |{\bf x}-{\bf x}_a|^7 }
\een
and set
\ben
T= { 8 \over 9} a^8\,, \qquad a= \left( { t+t_0 \over t_0} \right)   ^{ 1\over 9},
\een
\ben
x^0= T \cosh\left({ 8 z \over 9 t_0}\right)\,, \qquad x^{10}  =T \sinh\left({8z \over
  9t_0}\right),
\een
\ben
x^{\pm}= x^0 \pm x^{10}\, 
\een
with
\ben
dz + {dt \over a^9} =t_0 { 8 \over 9} {d x^+ \over x^+}
\een
we have
\ben
\fbox{$\displaystyle ds_{11}= d {\bf x}^2 - dx^+ dx ^- +
 \hat H \left({dx^+ \over x^+} \right)^2\,. $}    
\een

Note that 

\medskip 
\noindent $\bullet$This is a pp-wave whose profile depends on
light-cone time $x^+$.

\medskip 
\noindent $\bullet$ The solution is nevertheless {\sl
  boost-invariant},
the scalings  
\ben
x^+ \rightarrow \lambda x^+ \,,\qquad x^- \rightarrow \lambda ^{-1}  x^-
\een
with $\lambda \in {\Bbb R} \setminus 0$, leave the metric invariant.

\medskip 
\noindent $\bullet$ Reduction on the boost Killing vector  gives the
10-dimensional solution.

\medskip \noindent $\bullet$ The 11-dimensional solution is BPS, it admits a
covariantly constant Killing spinor but this is {\sl not} invariant
under boost and hence the 10-dimensional solution is not BPS.

\section {Conclusions}
\medskip 
\noindent $\bullet$ Homothetic solutions of classical matrix theory
resemble expanding universes but do not really capture the cosmology of
D0-branes.

\medskip 
\noindent $\bullet$ Exact supergravity 10-dimensional Type IIA
 solutions for expanding
universes
of D0-particles are available.

\medskip 
\noindent $\bullet$ Lifted to 11-dimensions they are vacuum 
pp-wave solutions with time dependent profile and hence BPS. 
Their reduction to 10-dimensions is on a boost Killing field
and hence they are time dependent and non-BPS in 10 dimensions.

\medskip 
\noindent $\bullet$ It seems that Quantum Mechanical matrix theory 
in a suitable limit captures the behaviour of the classical
super-gravity solutions.

\medskip 
\noindent $\bullet$ The status of the `Wave function of the
Universe'  remains  unclear. 

\section{Acknowledgements}

The authors thank S. Alexander, A. Guth, J. Khoury, Y. Okawa,
M. Sheikh-Jabbari, S. Shenker, and W. Taylor for useful
discussions. This research was supported by DOE contract
\#DE-FC02-94ER40818 and by NSF grant PHY-00-96515.

\end{document}